\documentclass[a4paper,10pt]{article}

\usepackage{epsfig}
\usepackage{amsfonts}
\usepackage{amsmath}
\usepackage{color}

\newcommand{\be}{\begin{equation}}
\newcommand{\ee}{\end{equation}}
\newcommand{\beq}{\begin{equation}}
\newcommand{\eeq}{\end{equation}}
\newcommand{\bea}{\begin{eqnarray}}
\newcommand{\eea}{\end{eqnarray}}
\newcommand{\nn}{\nonumber \\}

\newcommand{\bx}{{\bf x}}
\newcommand{\bp}{{\bf p}}

\newcommand{\br}{{\bf r}}
\newcommand{\bk}{{\bf k}}

\newcommand{\bs}{{\bf s}}

\newcommand{\NB}{n_{\mathrm{B}}}
\newcommand{\NF}{n_{\mathrm{F}}}
\newcommand{\eps}{{\varepsilon}}

\newcommand{\CF}{C_{\rm{F}}}
\newcommand{\Nc}{N_{\rm c}}

\def\lsi{\raise0.3ex\hbox{$<$\kern-0.75em\raise-1.1ex\hbox{$\sim$}}}
\def\gsi{\raise0.3ex\hbox{$>$\kern-0.75em\raise-1.1ex\hbox{$\sim$}}}

\newcommand{\disc}{\mathop{\mbox{Disc\,}}}

 \newcommand{\nF}[1]{n_\rmii{F{#1}}}
 \newcommand{\nB}[1]{n_\rmii{B{#1}}}
\newcommand{\rmii}[1]{{\mbox{\tiny\rm{#1}}}}

\newcommand{\Tint}[1]{{\hbox{$\sum$}\!\!\!\!\!\!\!\int\,}_{\!\!\!\!\raise-0.9ex\hbox{$\scriptstyle{#1}$}}}
\newcommand{\Tinti}[1]{{{\Sigma}\!\!\!\!\raise0.3ex\hbox{$\int$}_\rmii{${#1}$}}}

\newcommand{\bi}{\begin{itemize}}
\newcommand{\ei}{\end{itemize}}

\newcommand{\hide}[1]{ }

\title{Quarkonium spectral function in medium at next-to-leading order for any quark mass}
\author{Yannis Burnier\\{\it\small Institute of Theoretical Physics, EPFL, 
        CH-1015 Lausanne, Switzerland}}

\begin{document}

\maketitle

\begin{abstract}
The vector channel spectral function at zero spatial momentum is calculated at next-to-leading order in thermal QCD for any quark mass. It corresponds to the imaginary part of the massive quark contribution to the photon polarization tensor. The spectrum shows a well defined transport peak in contrast to both the heavy quark limit studied previously, where the low frequency domain is exponentially suppressed at this order and the naive massless case where it vanishes at leading order and diverges at next-to-leading order. 
From our general expressions, the massless limit can be taken and we show that no divergences occur if done carefully. Finally, we compare the massless limit to results from lattice simulations.
\end{abstract}

\section{Introduction}
In heavy ion collisions, fireballs of quark-gluon plasma are formed. The QCD matter being strongly interacting, the quarks and gluons only escape as mesons or hadrons when the plasma cooled down sufficiently. To reconstruct what happened at early stages of the collision, we have to resort to probes that can be traced in experiment and whose properties are modified in the plasma. For several available probes, one quantity describes their fate in the plasma: the vector channel spectral function in medium. 

For instance the way bound states such as charmonium or bottomonium decay into muon pairs \cite{McLerran:1984ay, Weldon:1990iw} could tell us how they are affected by the plasma \cite{Shuryak:1978ij,Matsui:1986dk}. Another example is the heavy flavor diffusion coefficient, which describes how the massive quarks will be diffused or slowed down by the plasma. This is another observable which is of interest for experiment, in fact the heavy quarks can be tagged and their transverse momentum distribution studied (see for instance \cite{Abelev:2014ipa, Adamczyk:2014uip}. For light quarks, the zero frequency limit of the spectral function describes the electric conductivity \cite{Kapusta} and its momentum depencance the photon or dilepton emission rate \cite{Aarts:2005hg, Laine:2013vma, Ghisoiu:2014mha}.

In the vacuum, this spectral function is known at five loops \cite{Baikov:2009uw} for massless fermions and Taylor expansions in the mass are known to four loops \cite{Hoang:2008qy,Kallen:1955fb,Barbieri:1973mt,Broadhurst:1993mw}. In the presence of a finite temperature medium, the two loop massless result is known since a long time \cite{Altherr:1989jc} and the case of large masses with respect to the temperature $M\gg T$ was calculated rather recently \cite{Burnier:2008ia}. Here we extend the previous calculations to any mass and discuss the transport part of the spectrum, which is suppressed in the heavy quark limit and was not obtained in the previous calculations. We still consider implicitly that the frequency $\omega\gg gT $ is sufficiently large so that we can neglect hard thermal loop (HTL) corrections \cite{Braaten:1990wp} in the fermion propagators and vertices. Other HTL corrections are of higher order \cite{Burnier:2008ia} and will not be addressed here either.

Of course in QCD, the convergence of perturbation theory is slow due to the largeness of the coupling $\alpha_s$ and moreover finite temperature gauge theories suffer from infrared problems 
so that the full infrared physics requires nonperturbative methods.
Lattice computations contain the full physics but are performed in Euclidean time and do not have a direct access to the Minkowskian spectral function. After measuring the corresponding Euclidean correlator, an analytic continuation is needed to obtain the desired spectral function. In the case of discrete numerical data of finite precision the reconstruction of the spectrum is very hard to perform \cite{Asakawa:2000tr,Rothkopf:2011ef,Burnier:2013nla,Burnier:2011jq}. The challenge is even bigger here since the Euclidean correlator is not even continuous at zero Euclidean times and hence the full analytical continuation is ill defined \cite{cuniberti}. That's where perturbation theory could again be of use, since the zero Euclidean time limit (or the corresponding large frequency limit of the spectral function) is weakly coupled and accessible to perturbation theory. This 'large' perturbative part (containing zero and possibly finite temperature contributions) could be subtracted from the lattice data \cite{Burnier:2012ts} or used as a prior to define the analytical continuation \cite{Ding:2012sp}. 

In the case of the vector current spectral function considered here, the Euclidean corralator was computed recently together with its mass dependence in \cite{Burnier:2012ze}.
In this paper we complete our program and calculate the related spectral function i.e. perform the analytic continuation. This is not a trivial task even though the Euclidean correlator $G(\tau,\bp)$ of ref.~\cite{Burnier:2012ze} can be cross checked by convoluting the spectral function $\rho(\omega,\bp)$ with the finite temperature kernel $K(\tau,\omega)$:
\beq
G(\tau,\bp)=\int_{0}^{\infty} \frac{d\omega}{ \pi}\;\rho(\omega,\bp)K(\tau,\omega),\quad K(\tau,\omega)=\frac{\cosh(\omega(\tau-\beta/2))}{\sinh(\omega\beta/2)}.\label{G}
\eeq

After defining the observables in section  \ref{sec:2}, we discuss the calculation in sec \ref{sec:3}. and refer to appendices for details. In section \ref{sec:4} we present our results for the spectrum, discuss the transport coefficients and derive the massless limit, which we compare to lattice results. Conclusions are given in section \ref{sec:6}.

\section{Correlators and spectral functions}\label{sec:2}

\subsection{Basic setup}
We consider the vector current of a massive quark described by the operator $\Psi(\tau,\bx)$
\beq 
J_\mu(\tau,\bx)=\bar \Psi(\tau,\bx)\gamma_\mu\Psi(\tau,\bx),
\eeq
with $\mu=0,..,d$. The object we compute here is the spectral function in medium, which is given by the thermal average of the current commutator
\beq
\rho^V(\omega)=\int dt\; e^{i\omega t}\int d^{d}x \langle \frac12 \left[J^\mu(t,x),J_\mu(0,0)\right]\rangle_T.
\eeq
Following \cite{Burnier:2008ia} we will start from the Euclidean correlator in frequency space ($\omega_n=2\pi n T$ denote the Matsubara frequencies),
\beq
G_E(\omega_n)=\int_0^\beta d\tau e^{i\omega_n \tau}\int d^{d}x\langle J^\mu(\tau,x)J_\mu(0,0)\rangle,
\eeq
which can be calculated using conventional finite temperature Feynman rules \cite{Kapusta,Lebellac,Norton:1974bm}. The spectral function can be determined from the discontinuity of the Euclidean correlator along the imaginary axis:
\beq
\rho^V(\omega)=\disc[G_E(-i\omega)]=\frac{1}{2i}\lim_{\eps\to 0^+}[G_E(-i\omega+\eps)-G_E(-i\omega-\eps)].\label{disc}
\eeq
Note that for $\omega\sim gT$ usual perturbation theory breaks down and would require to use Hard Thermal Loop (HTL) resummation, which will not be considered here. As was shown in \cite{Burnier:2008ia}, no infrared divergences occur so that HTL corrections are subleading for $\omega\gg gT$.

\subsection{Possible applications}

One observable defined by this spectral function is the production rate of muon pairs from the decay of massive quark pairs \cite{McLerran:1984ay, Weldon:1990iw}. If we suppose that the quark pair is at rest we have in particular:
\beq
\frac{dN_{\mu\bar\mu}}{d^4x d^4q}=\frac{-2e^4Z^2}{3(2\pi)^5 \omega^2}\left(1+\frac{2m^2_\mu}{\omega^2}\right)\left(1-\frac{4m^2_\mu}{\omega^2}\right)^{\frac12}n_B(\omega)\rho^V(\omega),
\eeq
where $Ze$ is the charge of the quark and $n_B$ the Bose-Einstein Distribution and $\omega=E_{\mu^+}+E_{\mu^-}$. The main contribution to this observable comes form the threshold $\omega\sim 2M$, where the spectrum can be obtained more precisely with dedicated resummations \cite{Burnier:2008ia,Burnier:2007qm}.

When speaking of heavy flavor diffusion or electric conductivity, the diffusion coefficient $D$ is obtained from the low energy behavior of the spectrum. In fact one expects the spectral function to look like a Lorentzian in the low energy limit
\beq
-\frac{\rho^V(\omega)}{\omega}\overset{0<\omega<\omega_{UV}}{\approx}3 \chi D\frac{\eta_D^2}{\eta_{D}^2+\omega^2},\label{rho_lowfrequ}
\eeq
where $\chi$ is the susceptibility, $\eta_D$ another number called the drag coefficient and $\omega_{UV}$ the scale at which other kind of physics enter and where the spectrum deviates from a Lorentzian\footnote{Note also that $\rho^V(\omega>0)<0$ hence the minus sign.}.
The diffusion coefficient can then be extracted as
\beq
D=-\frac{1}{3\chi}\lim_{\omega\to 0}\frac{\rho^V(\omega)}{\omega}.\label{Dc}
\eeq
For a thorough discussion of this formula and the zero frequency limit see for instance ref.~\cite{Kapusta}. 

If the onset of non-transport physics $\omega_{UV}$ is well separated from the transport peak, another strategy can be used \cite{CasalderreySolana:2006rq}. The idea is to calculate another observable, the momentum diffusion coefficient $\kappa$, which is proportional to the drag coefficient $\eta_D$. It can be extracted from the falloff of the Lorentzian peak \cite{CaronHuot:2009uh}:
\beq
\kappa=2 M_{kin} T \eta_D\approx \left.-\frac{2M_{kin}^2\omega^2}{3\chi}\frac{\rho^V(\omega)}{\omega} \right|_{\eta_D\ll \omega \ll\omega_{UV}} \label{kappa}\, ,
\eeq
where $M_{kin}$ is the in medium kinetic mass defined by the low momentum limit of the dispersion relation.\footnote{Namely the velocity dependence of the free energy is expanded as $F(v)=M_{rest}+M_{kin}v^2/2+\mathcal{O}(v^4)$.}
The fluctuation-dissipation theorem finally relates this coefficient to the diffusion coefficient $D=2 T^2/\kappa$ and hence to the drag coefficient $\eta_D=\kappa/(2M_{kin} T)$.

The momentum diffusion coefficient can be calculated in perturbation theory with dedicated resummations \cite{Moore:2004tg}. However even if the resummations seem to catch the relevant physics, the convergence of the perturbative series for $\kappa$ is at best very slow. In the case of the heavy quarks for instance, the first non-vanishing contribution arises at $\mathcal{O}(\alpha^2)$ and the correction $\mathcal{O}(\alpha^2g)$ is an order of magnitude larger for typical heavy ion plasmas \cite{CaronHuot:2007gq}. 

\section{Outline of the calculation}\label{sec:3}

We now turn to the calculation of the spectral function, following refs.~\cite{Burnier:2008ia,Burnier:2012ze,Burnier:2007qm,Burnier:2013vsa}.

\subsection{Leading order and notations}

Performing the Wick contractions and the trace algebra, we get at leading order (LO):
\beq
G_E^{V}=2\Nc\Tint {\{P\}}\biggl(-\frac{4 (1-\epsilon )}{\Delta (P)}+\frac{-4 M^2+2 Q^2
   (1-\epsilon )}{\Delta (P) \Delta (P-Q)}\biggr)\nn
\eeq
so that 
\bea
\rho^V_{LO}(\omega)&=&\disc[G^V_E(-i \omega)]\\ \notag&=&2\Nc\disc[\Tint {\{P\}}\biggl(-\frac{4 (1-\epsilon )}{\Delta (P)}+\frac{-4 M^2+2 Q^2
   (1-\epsilon )}{\Delta (P) \Delta (P-Q)}\biggr)],
\eea
where $Q=(-i\omega\pm \epsilon,\bf 0)$ will be set in the process of taking the discontinuity. Note that the first term is independent of $Q$ and will not contribute to the spectrum. 
To simplify the following expressions in both the LO and the next-to-leading order (NLO), we introduce the following notations:
\bea
I_{i j}&=&\disc\left[\Tint {\{P\}}\frac{1}{\Delta (P)^i\Delta(P-Q)^j}\right],\label{defI}\\
I_{ijklm}^n&=&\disc\left[\Tint {K,\{P\}}\frac{(K\cdot Q)^n}{(K^2)^i\Delta (P)^j\Delta(P-K)^k\Delta(P-Q)^l\Delta(P-K-Q)^m}\right]\notag
\eea
with $\Delta(P)=P^2-M^2$, so that at leading order,
\beq
\rho^V_{LO}(\omega)=-2\Nc(4 M^2+2 \omega^2)I_{11}(\omega).\label{LO}
\eeq
All the relevant I's are calculated in appendix \ref{I's} and in particular:
\bea 
I_{11}(\omega)&=\notag&\theta(\omega-2M)\frac{\sqrt{\omega^2-4M^2}}{16\pi \omega}\tanh\left(\frac{\omega}{4T}\right)\left[1+\eps\left(2+\ln\frac{\bar\mu^2}{\omega^2-4 M^2}\right)\right]\\&&+\pi \omega\delta(\omega)\int_p \frac{\NF'(E_p)}{2E_p^2},\label{LO2}
\eea
where we introduced the notation $E_p^2={\bf p}^2+M^2$, $\int_p=\int \frac{d^dp}{(2 \pi)^d}$.

\subsection{Next to leading order}
After performing the Wick contractions and the trace algebra, we get the NLO in terms of master sum-integrals defined in equations (\ref{defI}):
\bea
\frac{\rho^{V}_{NLO}}{4 \Nc\CF g^2}
   &=&4 (1-\epsilon)^2I_{12000}^0 -4 (1-\epsilon)I_{11100}^0+8 (1-\epsilon)M^2 I_{12100}^0\notag\\[-6pt]&&-4 (1-\epsilon)^2 I_{02100}^0
   -8 (1-\epsilon)^2 I^1_{11110}
   +8(1-\epsilon)  I^0_{01110}\nn&&
   -8(2 M^2+\omega^2(1-\epsilon
   ))I^0_{11110}+4(1-\epsilon)I_{10110}^0\nn&&
   +8M^2 \left(2 M^2+\omega^2
   (1-\epsilon)\right) I_{12110}^0\nn&&
   -4(1-\epsilon) \left(2 M^2+\omega^2
   (1-\epsilon)\right)I_{02110}^0\nn&&-2(1-\epsilon)I_{-11111}^0+2 \left(2 M^2 \epsilon +\omega^2
   \left(2-\epsilon-\epsilon^2\right)\right)I_{01111}^0 \nn&&+2 \left(4 M^4-2 M^2 \omega^2 \epsilon -\omega^4 (1-\epsilon
   )\right)I_{11111}^0\nn&&
   -4(1-\epsilon)I_{11010}^0+
4 (1-\epsilon) \left(2
   M^2+\omega^2 (1-\epsilon)\right)I^0_{12010}.\label{NLO}\eea
Note that the first four terms are independent of $\omega$ and do not contribute to the spectral function.

\subsection{Renormalization}

The previous NLO expression (\ref{NLO}) is UV divergent but is finite after redefinition of the mass.
The counterterms for the currents read:
\beq
\frac{\rho^{V,CT}_{NLO}}{4\Nc \CF g^2}=
\frac{\delta M^2}{g^2\CF}\frac{ 1}{4\Nc}\frac{ \partial \rho^V_{LO}}{\partial M^2}\label{CT}
\eeq
with
\beq
\frac{ 1}{4\Nc}\frac{ \partial \rho^V_{LO}}{\partial M^2}=(4 M^2+2 \omega^2 (1-\epsilon)) I_{21}-2I_{11}-2(1-\epsilon) I_{20}
\eeq
and using the pole mass scheme as in ref.~\cite{Burnier:2012ze}, we set
\bea
\delta M^2&=&-\frac{6g^2\CF M^2}{(4\pi)^2}\left(\frac{1}{\epsilon}+\ln\frac{\bar\mu^2}{M^2}+\frac43 \right)
\notag\\&=&-g^2\CF\int_{k} \frac{(2-2\epsilon) (E_{pk}-k)+\frac{2 M^2}{\Delta_{-+}}+\frac{2
   M^2}{\Delta_{++}}}{2 k E_{pk}},
\eea
where we denoted $E_{pk}=E_{\bf p+\bf k}$, $k=|\bf k|$ and $\Delta_{\pm\pm}=k\pm E_p\pm E_{pk}$.

\subsection{Thermal correction to the mass}
After subtraction of the counterterms, the spectral function is finite everywhere but at the threshold $\omega=2M$. The divergence there comes from thermal corrections, in fact one can rewrite the whole spectral function as
\beq
\rho^V(\omega,M^2)=\rho^V_{LO}(\omega,M^2)+\delta M_T^2 \frac{\partial\rho^V_{LO}(\omega,M^2)}{\partial M^2}+\bar\rho^V_{NLO}(\omega,M^2)+\mathcal{O}(g^4),
\eeq
where 
\beq
\bar\rho^V_{NLO}=\rho^V_{NLO}-\delta M_T^2 \frac{\partial\rho^V_{LO}}{\partial M^2}\label{MS}
\eeq
and the term $\delta M_T^2 \frac{\partial\rho^V_{LO}(\omega,M^2)}{\partial M^2}$ is actually responsible for the divergence. 
In fact we can resum this contribution by redefining the mass $M^2\to M^2+\delta M^2_T$:
\beq
\rho^V(\omega,M^2)=\rho^V_{LO}(\omega,M^2+\delta M_T^2)+\bar\rho^V_{NLO}(\omega,M^2+\delta M_T^2)+\mathcal{O}(g^4).\label{resumM}
\eeq
The explicit shift $\delta M^2_T$, is the thermal contribution to the dispersion relation, which, for a massless fermion is the well-known
\beq 
\delta M^2_T=g^2 \CF \int_0^\infty \frac{dk}{\pi^2} k(\NB(k)+\NF(k))=\frac{g^2 \CF T^2}{4} \label{MassShift}
\eeq 
and for a massive fermion, the less well known \cite{Seibert:1993aw} expression
\bea
\frac{\delta M^2_T}{g^2\CF}&=&2\int_k\frac{\NB(k)}{k}+\frac{\NF(E_{pk})}{ E_{pk}}\left(1-\frac{M^2}{\Delta_{++}\Delta_{--}}-\frac{M^2}{\Delta_{+-}\Delta_{-+}}\right)\\&=&\int_0^\infty \frac{dk}{\pi^2} \left(k\,\NB(k)+\frac{k^2}{E_k}\left(1+\frac{M^2}{2 p k} \ln\left|\frac{k-p}{k+p}\right|\right)\NF(E_k)\right),\notag
\eea
which actually depends on the integration variable $p$.

As a summary, to avoid divergences at the threshold, we resum the mass correction. To calculate the spectral function of a fermion of mass squared $M^2$, we calculate the spectral function at the mass $M^2+\delta M^2$ as written in equation (\ref{resumM}) and modify the NLO contribution as explained in equation (\ref{MS}). 

Note that the relevant bosonic part of this shift was performed in \cite{Burnier:2008ia}. The divergence is however integrable and the shift was left out in 
ref.~\cite{Burnier:2012ze}, where the Euclidean correlator was calculated, but the changes are easily tractable (see appendix \ref{A4}). 

\subsection{Explicit calculation of the NLO result}

While the leading order is given in equations (\ref{LO},\ref{LO2}) the next-to-leading order requires significantly more work. The full expression is obtained adding to the NLO (\ref{NLO}), the mass counterterm (\ref{CT}) and the contribution from the thermal mass shift (\ref{MS}). 
The first step is to carry out the sums (see Appendix~\ref{calc}) and take the discontinuity (\ref{disc}). We are then left with the integrals and a delta function remaining from the discontinuity. Parts of the integrals can be performed analytically and the remaining ones have to be done numerically.
The explicit expressions for the master integrals are given in appendix \ref{I's} and the details on their integration in appendix \ref{C}. 

\section{Results}\label{sec:4}

The final result can be split into three parts
\beq
\rho_{NLO}^V(\omega)=\rho_{NLO}^{vac}(\omega)+\rho_{NLO}^{bos}(\omega)+\rho_{NLO}^{ferm}(\omega).
\eeq
First the vacuum part \cite{Kallen:1955fb,Barbieri:1973mt,Broadhurst:1993mw} that can be integrated explicitly
\bea
\frac{\rho^{vac}_{NLO}(\omega)}{4\Nc \CF g^2}&=&\frac{2\theta(\omega-2M)}{(4\pi)^3 \omega^2} \label{rhoNLOvac}
\biggl[
(4 M^4-\omega^4)L_2\left(\frac{\omega-\sqrt{\omega^2-4 M^2}}
{\omega+\sqrt{\omega^2 - 4 M^2}}\right)
\\&&\notag+(7 M^4+2 M^2\omega^2-3\omega^4)
\mathrm{acosh}\left(\frac{\omega}{2 M}\right)
+\omega \sqrt{\omega^2-4 M^2}\\ \notag&&\times
\left((\omega^2+2M^2) \ln \frac{\omega(\omega^2-4 M^2)}{M^3}
 -\frac38(\omega^2+6 M^2) \right)\biggr],
\eea
where $L_2=4\mathrm{Li}_2(x)+2\mathrm{Li}_2(-x)+\ln(x)\left[2\ln(1-x)+\ln(1+x)\right]$.
Secondly the first thermal part, that we will denote 'bosonic' thermal correction, calculated in ref.~\cite{Burnier:2008ia} for which one integral is left for numerical evaluation (for mass shift contribution see Appendix \ref{A4}). It is proportional to the Bose-Einstein distribution function $\NB(k)$ and does not contain any Fermi-Dirac distribution:
\bea
&&\hspace{-5mm}\frac{\rho^{bos}_{NLO}}{4\Nc \CF g^2 } = \label{rhoNLObos}
\frac{2}{(4\pi)^3 \omega^2}
\int_0^\infty dk \, \frac{\NB(k)}{k}\biggl\{ \theta(\omega)
\theta\left(k-\frac{4M^2-\omega^2}{2\omega}\right)
\biggl[
\\&&\notag  2 \omega^2 k^2 \sqrt{1-\frac{4 M^2}{\omega(\omega+2 k)}}
+(\omega^2 + 2 M^2) \sqrt{\omega(\omega+2 k)}\sqrt{\omega(\omega+2 k)- 4M^2}
\\&&\notag- 2 \Bigl(\omega^4 - 4 M^4 + 2 \omega k (\omega^2 + 2 M^2) + 2 \omega^2 k^2\Bigr)
\mathrm{acosh} \sqrt{\frac{\omega(\omega+2k)}{4M^2}} \biggr]
\\&&\notag+\theta(\omega - 2M)\theta\Bigl(\frac{\omega^2-4M^2}{2\omega}-k\Bigr)\biggl[2 \omega^2 k^2 \sqrt{1-\frac{4 M^2}{\omega(\omega-2 k)}}
\\&&\notag +(\omega^2 + 2 M^2) \sqrt{\omega(\omega-2 k)}\sqrt{\omega(\omega-2 k)- 4M^2}
\\&&\notag- 2 \Bigl(\omega^4 - 4 M^4 - 2 \omega k (\omega^2 + 2 M^2) 
+ 2 \omega^2 k^2\Bigr)
\mathrm{acosh} \sqrt{\frac{\omega(\omega-2k)}{4M^2}} \biggr]
\\&&\notag + \theta(\omega - 2M) \biggl[ - 2 (\omega^2 + 2 M^2)\, \omega \sqrt{\omega^2 - 4 M^2}
\\&&\notag + 4 \Bigl(\omega^4 - 4 M^4 + 2 \omega^2 k^2\Bigr)\mathrm{acosh} \biggl( \frac{\omega}{2M} \biggr) 
  \biggr] \biggr\}.
\eea 
The remaining contribution $\rho^{ferm}_{NLO}$ contains at least one Fermi-Dirac distribution $\NF(E_p)$ or $ \NF(E_{pk})$ hence it is suppressed in the limit $M\ll T$. However it is the only piece containing a non-vanishing transport peak near $\omega\to 0$ so that it dominates the spectrum at low frequency. In this part, two integrals have to be performed numerically. The full expression is rather lengthy and can be read from appendix \ref{C}. For $\omega<2M$ only a few structures actually contribute and we can quote the result here: 
\beq
\rho^{ferm}_{NLO}\overset{\omega<2M}{=} \label{rhoNLOferm}
-2\NF\left(\frac{\omega}{2}\right)\rho^{bos}_{NLO}+\rho^{ferm,1}_{NLO},
\eeq
where
\bea
 &&\hspace{-5mm}\frac{\rho^{ferm,1}_{NLO}}{4\Nc \CF g^2}\label{rhoNLOferm1}=\frac{1}{8 \pi^3}\int_{\omega/2}^\infty \int_{E_p^{1+}}^\infty
\Biggl\{ \\&&\notag 1+\frac{M^2 \left(2 M^2+\omega ^2\right)}{\omega
   ^2} \left(\frac{1}{(2 E_p+2 k-\omega )^2}+\frac{1}{(2 E_p-\omega )^2}\right)\\&&\notag+\frac{-2 k^2 \omega ^2+\left(k \omega +2 M^2\right)^2-\omega ^2 (\omega
   -k)^2}{2 k \omega ^2}\left(\frac{1}{2 E_p+2 k-\omega }-\frac{1}{2 E_p-\omega }\right)\Biggr\} \\\notag &&\times [\NB(k) (\NF(E_p)-\NF(E_p+k-\omega ))+(\NF(E_p)-1) \NF(E_p+k-\omega )].
\eea
This last term $\rho^{ferm,1}_{NLO}$ contribute to the transport peak and comes form gluon emission or absorption by the massive fermion. 

\begin{figure}
\begin{center}
\includegraphics[height=4.9cm]{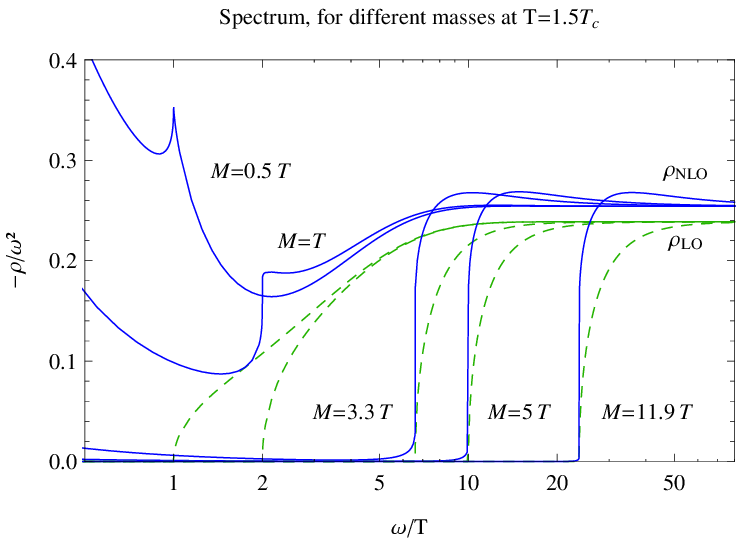}
\includegraphics[height=4.9cm]{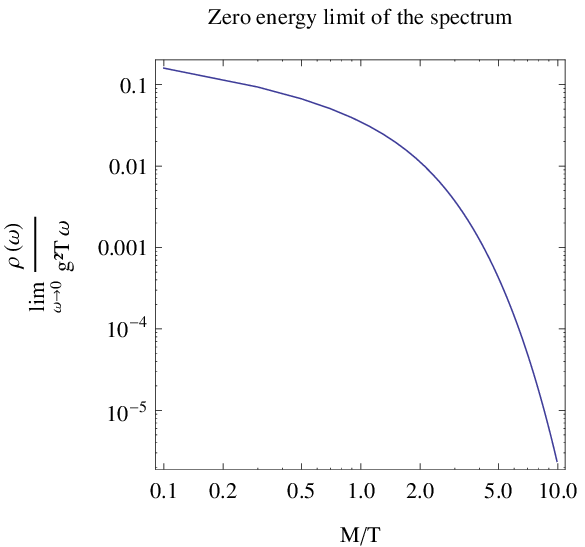}
\caption{(Left): LO (green dashed) and NLO (blue) spectral function ($-\rho^V(\omega)/\omega^2$) for $M=(0.5,1,3.3,5,10)T$, $T=1.5 T_c$. The LO vanishes at frequencies below the threshold $\omega=2M$ where the NLO shows a discontinuity. (Right) Value of the low energy limit of $\rho^V(\omega)/\omega$ as function of the quark mass.}\label{fig:0} \vspace{-0.4cm}
\end{center}
\end{figure}

The full result containing the LO and NLO contributions is plotted in fig.~\ref{fig:0} (left) for $T=1.5 T_c$ and different masses ranging from $M=0.5T$ to $M=11.9T$ corresponding to a bottom quark of mass 4.65GeV. Note that the spectral function is in fact negative and for clarity we show $-\rho^V(\omega)$ in the plots. In fig.~\ref{fig:0} (right) we show the zero frequency limit of the spectral function $\lim_{\omega\to0}\rho^V(\omega)/\omega$, which enters the determination of the transport coefficient $D$. The transport coefficient itself requires division by the suscptibility $\chi$, for which we use the NLO result of ref.~\cite{Burnier:2012ze}, see fig.~\ref{fig:NLO}.
Separate results for the three contributions to the next to leading order are shown in fig.~\ref{fig:NLO} for the case $M=3.3T$, representing the charm quark at $T=1.5 T_c$ studied on the lattice in \cite{Ding:2010yz} and for the generic case $M=T$ in fig.~\ref{M=1}. In these figures we scaled the result to $\omega^2$ so that the vacuum contribution goes to a constant at large frequency. 

\subsection{Charm transport}

\begin{figure}
\includegraphics[height=4.9cm]{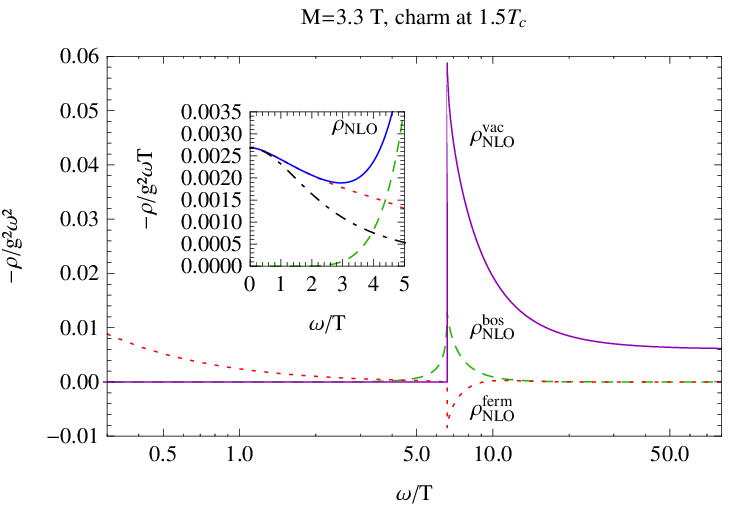}
\includegraphics[height=4.9cm]{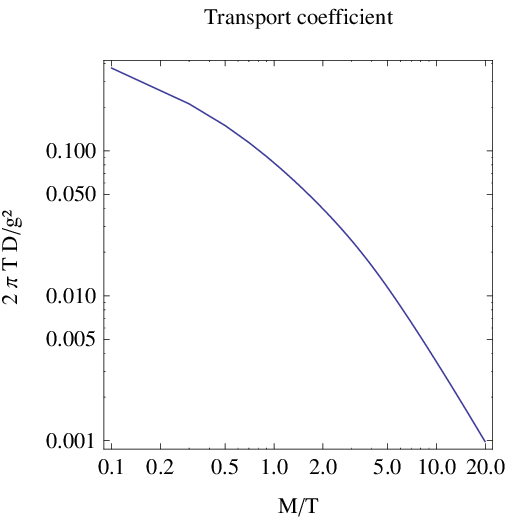}
\caption{(Left): Different parts of the NLO contribution to the spectral function for M=$3.3T$ and Lorentzian fit of the transport peak (black dot-dashed line) (Right): The generated transport coefficient $D$.}\label{fig:NLO} \vspace{-0.4cm}
\end{figure}

In the insert of the same figures \ref{fig:NLO}, \ref{M=1}, we scale the spectrum to the frequency and zoom on the low frequency region to see the transport peak. We see that $\frac{\rho(\omega)}{\omega}$ goes smoothly to a constant with vanishing slope at $\omega\to 0$ and typical Lorentzian curvature, hence defining a diffusion coefficient $D$ according to equation (\ref{Dc}). The value of $D$ is plotted as a function of the quark mass in fig. \ref{fig:NLO}(right). We see that it is suppressed for large masses $T D\propto (M/T)^{-2}$ and behaves as a power law $T D\propto (M/T)^{-\frac14}$ at small $M$. In the case of the charm quark at $T=1.5T_c$, we see that it is in principle small $2\pi T D\approx 0.1$ in comparison to the lattice results suggesting $2\pi T D\approx 2$ \cite{Ding:2010yz} and to the perturbative heavy quark limit $2\pi T D\approx 10$ \cite{CaronHuot:2007gq}.

That said, the transport peak is not well separated from the UV physics - at least in this low order calculation - and there is no region where $\eta_D\ll \omega \ll \omega_{UV}$ so that the momentum diffusion coefficient cannot be defined straight from formula (\ref{kappa}). Of course for the only purpose of defining $\kappa$ or $\eta_D$ one could just fit the low frequency part with equation (\ref{rho_lowfrequ}) and get $\eta_D$ as a fit parameter adjusted to the curvature of $\frac{\rho(\omega)}{\omega}$ around $\omega=0$. In the previous case we get $\eta_D\sim 2.5$ which translates, if one would still trust the fluctuation-dissipation theorem and supposing that $M_{kin}=M$, into $2 \pi T D\sim 0.8$. This differ from the direct estimate made above ($2\pi T D\approx 0.1$) showing that the fluctuation-dissipation theorem does not seem to apply here. Note again that the present computation is not systematic \cite{Moore:2006qn} for $\omega<gT$ so that no strong conclusion should be made with this remark.

\subsection{Massless limit and electric conductivity}

\begin{figure}
\includegraphics[height=4.8cm]{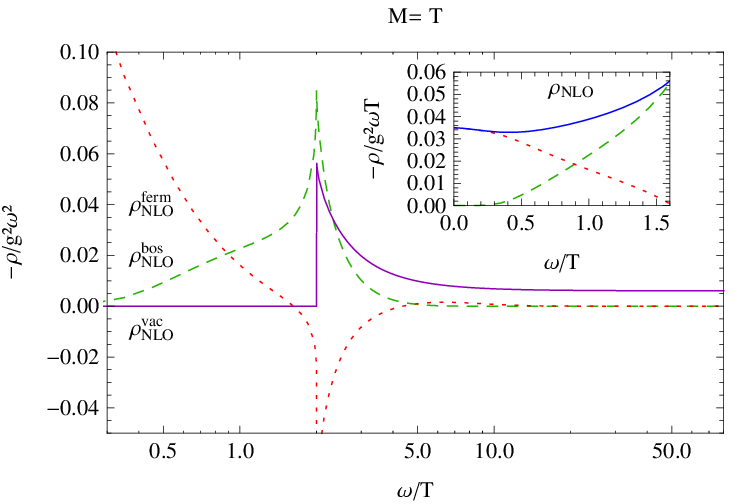}
\includegraphics[height=4.8cm]{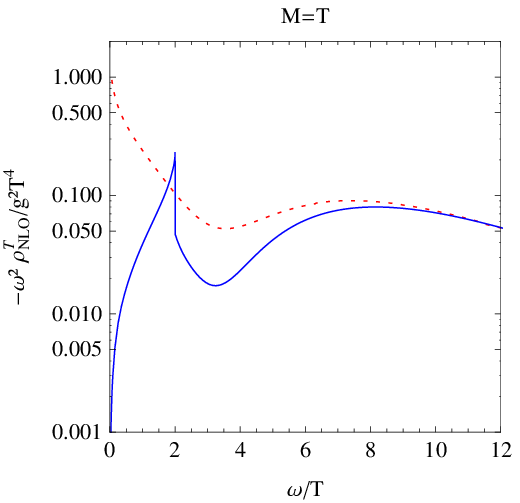}
\caption{(Left): The different contributions to the NLO spectrum for $M=T$. (Right): The result of this paper for the total thermal NLO contribution (continuous blue line) compared to the result of ref.~\cite{Altherr:1989jc} where the mass shift was not fully taken into account (dotted red line). Note that the threshold structure appears negative as we show only the thermal part of the NLO. For the complete spectrum see fig.~\ref{fig:0}}\label{M=1} \vspace{-0.4cm}
\end{figure}

Let's we consider a fermion that is massless in the vacuum. Even if in this case HTL correction would be of the same order as our NLO result, it is interesting to see how our result behaves. As we resummed the thermal mass correction according to equation (\ref{resumM}), we in fact have to calculate the spectral function of a fermion of mass $\delta M_T^2$. In a typical quark-gluon plasma produced in heavy ion collision we have numerically (\ref{MassShift}) $\delta M_T^2=g^2 \CF T^2/4 \approx T^2$. The spectrum of such a fermion is shown in fig. \ref{M=1}. We see that the spectrum has no divergence neither at the threshold nor at zero frequency and shows a transport peak. This result differs from the old result of ref.~\cite{Altherr:1989jc}. There, the mass shift was noticed and performed in the LO result but was not made in the NLO calculation where the mass was set to zero. Their NLO contribution hence diverges at zero frequency, whereas ours has a threshold structure at $\omega=2\delta M_T$ and a transport peak, see fig.~\ref{M=1}. At high frequency both results agree and merge to the asymptotic behavior derived in ref.~\cite{CaronHuot:2009ns}, which reads\footnote{The asymptotic behavior of ref.~\cite{CaronHuot:2009ns} was derived without the mass shift hence contains only the first term, see also footnote 6 there}
\beq
-\rho^T_{NLO}=-\rho^{bos}_{NLO}-\rho^{ferm}_{NLO}\overset{\omega\gg M,T}{=}4 \Nc \CF g^2\left(\frac{\pi T^4}{36\omega^2}+\frac{ T^2 M^2}{8 \pi\omega^2}\right).
\eeq

The transport coefficient obtained here is of order $2 \pi T D\sim 0.3$, which is again on the low side, the perturbative resummation from \cite{Arnold:2000dr,Arnold:2003zc} gets in this case $2 \pi T D\sim 25$ and lattice results ranges from $2 \pi T D\sim 1-6$ depending on the analytic continuation method used \cite{Ding:2010ga,Burnier:2011jq}.

\subsection{Matching the massless limit with lattice results}\label{sec:5}

\begin{figure}
\includegraphics[height=4cm]{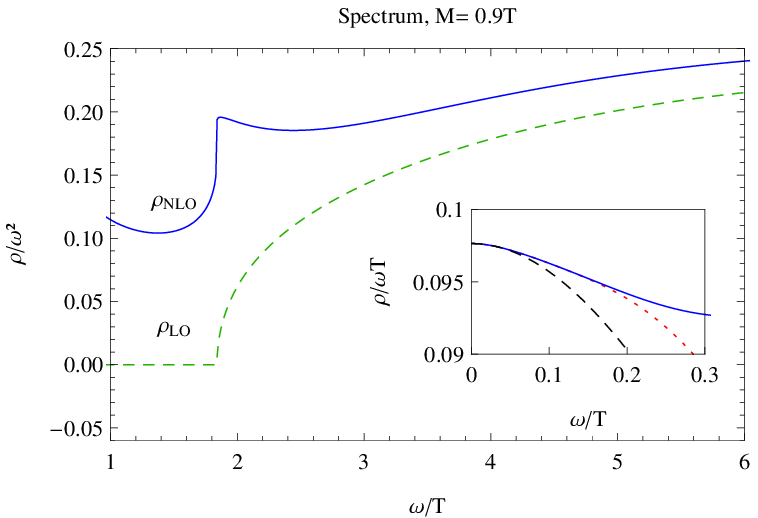}
\includegraphics[height=4cm]{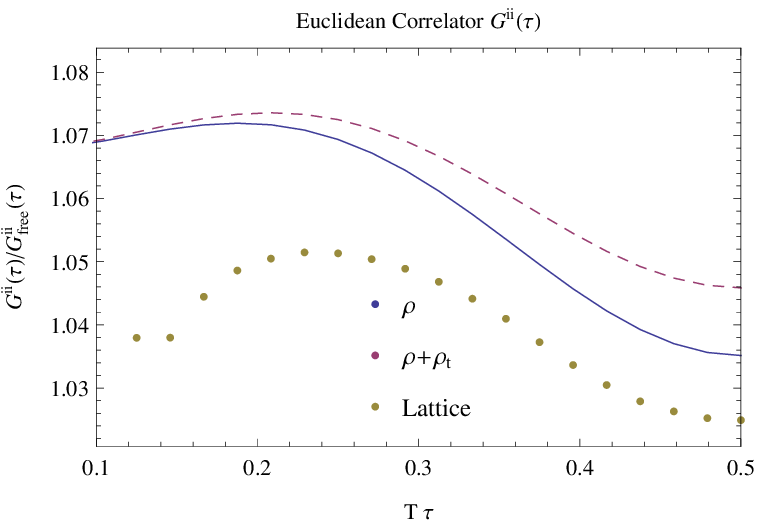}
\caption{(Left): Spectrum at leading and next-to-leading order for $M=0.9T$. In the insert, we fit the low energy spectrum with a Lorentzian. (Right): Next-to-leading order Euclidean correlator for $M=T$ (blue continuous line) together with lattice data (dots). We note that the agreement can be greatly improved adding more power to the transport peak, even if some normalization factor remains. (magenta dashed line)}\label{fig:spec_fct} \vspace{-0.4cm}
\end{figure}

The full spectral function for $M^2=\delta M_T^2=g^2\CF T/4$ with $T=1.46T_c$ is shown in fig.~\ref{fig:spec_fct} together with the Euclidean correlator scaled to the free correlator \cite{Aarts:2005hg}. Keeping in mind that our approximations are not consistent at low frequencies, we still compare the Euclidean correlator to the lattice data of ref. \cite{Ding:2010ga} for a massless fermion. 
Apart from an overall normalization\footnote{Note that the normalization of the perturbative curve could be improved by adding higher order in the vacuum spectral function.} we see a different slope at large $\tau$ signaling a lack of power in the small $\omega$ region. 
This could be compensated for by an additional transport peak. Keeping for instance the same width $\eta_D=0.7$, if we add an additional $-\rho_t/\omega=\frac{3 \bar D \chi \eta_D^2}{\omega^2+\eta_D^2}$ for $\omega<2\delta M_T$ with $2\pi T \bar D=0.05$ we get the dashed curve on fig.~\ref{fig:spec_fct}b). We see that it now goes nicely parallel to the lattice data. The total diffusion coefficient would then be $2 \pi T D\sim 0.4$, still on the low side.

\section{Conclusion}\label{sec:6}

We calculated the thermal correction to the massive quark vector spectral function at NLO in thermal QCD. The thermal corrections are small in comparison to the vacuum corrections for large fermion masses and comparable to them for $M\sim T$. 

The result shows some typical features: First, the threshold for pair production is smoothed by thermal corrections, the discontinuity in the vacuum spectrum being partly compensated by thermal effects. 
Secondly a transport peak appears at this order, it is however small in comparison to other expectations. We also see that the transport peak is broad and is not well separated from other kind of physics, rendering its determination via the fluctuation-dissipation theorem difficult. It should however be stressed that higher loop orders give corrections of the same order for $\omega\leq gT$.

Setting the vacuum fermion mass to zero in our formulas do not lead to divergences. This contradict the calculation of ref.~\cite{Altherr:1989jc} where the spectral function was calculated for massless fermions and the result for the NLO diverges at zero frequency hindering a definition of the corresponding Euclidean correlator. The difference can be traced back to how the thermal mass shift is introduced in the NLO. In the previous reference, the thermal mass shift was performed in the leading order part as done here but not in the NLO contribution where the fermion remained purely massless, leading to divergences at zero frequency.

\section*{Acknowledgements}

The author would like to thank M. Laine for helpful discussions and imporant comments on the manuscript. 
This work was supported by the SNF grant PZ00P2-142524.

\appendix
\section{Calculation of the master integrals}\label{calc}

The calculation of the master sum-integrals is mostly standard (for details see ref.~\cite{Norton:1974bm}), only one difficulty arises in double poles at zero frequency, which will be explained below. Their calculation is subtle and is shown explicitly for the case of $I_{01111}^0$, which is the simplest from the amount of algebra but contains most of the technical difficulties.

We start by rewriting the master integral in a form where the elementary summation formula (valid for
 $0<\tau<\beta$)
\bea
T\sum_{p_n}\frac{e^{i p_n \tau}}{p_n^2+E_p^2}&=&\frac{\NB(E_p)}{2 E_p}\left(e^{\tau E_p}+e^{(\beta-\tau)E_p}\right)\notag\\
T\sum_{\{p_n\}}\frac{e^{i p_n \tau}}{p_n^2+E_p^2}&=&\frac{\NF(E_p)}{2 E_p}\left(e^{\tau E_p}-e^{(\beta-\tau)E_p}\right)\label{sums}
\eea
can be applied. To do that we shift $P-K\to K$ and introduced the new integration variables $S,R$ and the corresponding delta functions:
\bea
I_{01111}^0&=&\disc\left[\Tint {K,\{P\}}\frac{1}{\Delta (P)\Delta(P-K)\Delta(P-Q)\Delta(P-K-Q)}\right]\notag\\&=&\disc\left[ \Tint {\{P,K,S,R\}}\frac{\delta^4(R-P+Q)\delta^4(S-K+Q)}{\Delta (P)\Delta(K)\Delta(R)\Delta(S)}\right].\label{Itosum}
\eea
The temporal part of the delta function can be rewritten as an integral, where, we keep track of time ordering:
\bea
&&\delta(r_n-p_n+q_n)\delta(s_n-k_n+q_n)\\&&=e^{i \beta(p_n+s_n)}\int_0^\beta d\tau_1 e^{i(r_n-p_n+q_n)\tau_1}\int_0^{\tau_1} d\tau_2 e^{i(-s_n+k_n-q_n)\tau_2}\\&&+e^{i \beta(p_n+s_n+q_n)}\int_0^\beta d\tau_1 e^{i(-s_n+k_n-q_n)\tau_1}\int_0^{\tau_1} d\tau_2 e^{i(r_n-p_n+q_n)\tau_2}\label{delta0}.
\eea
The factors $e^{i \beta(p_n+k_n)}=e^{i \beta(p_n+s_n+q_n)}=1$ were added so that the total phase of all the Matsubara frequencies are between $0$ and $\beta$. Thanks to the time ordering and this last prescription the sums can be performed with formula (\ref{sums}) and then the integrals over $\tau_1,\tau_2$ calculated. Remembering that $q_n$ is a bosonic Matusubara frequency we can set $e^{i q_n \beta}\to1$\footnote{This simplification has to be done, it is part of the prescription to get the correct analytic continuation.}. The result is a rather long expression containing many terms that can be simplified in rewriting all the exponents in terms of Fermi-Dirac (or Bose-Einstein) distributions. Each term contains a product of Fermi-Dirac (or Bose-Einstein) distributions in the numerator and products of all kinds of sums of the energies $(a_p E_p+a_k E_k+a_r E_r+a_s E_s+i a_q q_n)$ with integer $a_i$'s in the denominator. Note that there are no divergences (up to the poles for the $q_n$ variable on the complex axis) since when one sum of the energies $(a_p E_p+a_k E_k+a_r E_r+a_s E_s)$ vanishes in the denominator, the numerator vanishes as well. This cancellation obviously occurs since the initial integral $(\ref{delta0})$ has no poles.

Now, it is possible to extract the discontinuity in each of these terms. The difficulty arises in terms proportional to 
\beq
\frac{f(E_p,E_r,E_k,E_s,\omega)}{(\omega-E_p+E_r)(\omega+E_s-E_k)}.
\eeq
This term contains a simple pole at $\omega=0$ when we enforce one of the delta functions contained in (\ref{Itosum}). Note that its contribution is finite even after replacing the second delta function as the numerator $f$ vanishes when energies are set equal. However this term also contains a double pole when both deltas are set to zero. 
The simplest way to deal with this issue is to rewrite the delta functions in (\ref{Itosum}) as 
\bea
\delta^3(\br-\bp)\delta^3(\bk-\bs)\propto  \delta(E_r-E_p)\delta^2(\Omega_r-\Omega_p)\delta(E_s-E_k)\delta(\Omega_s-\Omega_k)\notag\\
\propto \delta(E_r-E_p)\delta^2(\Omega_r-\Omega_p)\delta(E_s-E_k-E_r+E_p)\delta(\Omega_s-\Omega_k)
\eea
and use the $\delta(E_s-E_k-E_r+E_p)$ to replace $E_s-E_k$ by $E_r-E_p$ (note that this replacement has to be done as a limit). After this, all the poles which contribute at $\omega\sim0$ are of the type $(\omega\pm(E_r-E_p))$. 
The discontinuity can be taken in each term separately using
\beq
\disc\left[\frac{1}{\omega-A}\right]=-\pi \delta(\omega-A)
\eeq
for simple poles and 
\beq
\disc\left[\frac{1}{(\omega-A)^2}\right]=\pi \delta'(\omega-A)
\eeq
for double poles.

The spatial integrals over $\bf r,\bf s$ can be performed using the remaining delta functions $\delta^3({\bf r-\bf p})\delta^3({\bf s - \bf k})$. They force us to perform the limit $E_r\to E_p$ and get as final result:
\bea
I^{0~ (\omega\sim 0)}_{01111}=\int_{p,k} \left\{\pi\omega\delta(\omega)\left(\frac{\nF1'}{4E_p^2 E_{pk}^3}(1-2\nF2)+\frac{\nF1'\nF2'}{4 E_p^2 E_{pk}^2}\right)\right\}.
\eea
 
\section{Master integrals}\label{I's}

Following the notations of ref.~\cite{Burnier:2008ia}, we detail here the different master integrals (I denote $E_p^2=p^2+M^2,E_k^2=k^2+M^2,~E_{p-k}^2=(p+k)^2+M^2$, $k$ being the gluon momentum). First the leading order sum-integrals:
\bea
I_{11}&=&\pi \int_p\left[ \omega\delta(\omega)\frac{\nF1'}{2E_p^2}+ \left(\delta(\omega-2E_p)-\delta(\omega+2E_p)\right)\frac{1-2\nF1}{4 E_p^2}\right], \\
I_{21}&=&
\pi\int_p(1-2\epsilon)\left[\omega\delta(\omega)\frac{\nF1'}{8p^2 E_p^2}+\left(\delta(\omega-2E_p)-\delta(\omega+2E_p)\right)\frac{1-2\nF1}{16 p^2 E_p^2 }\right].\notag
\eea
From them a few NLO sum-integrals are derived:
\bea
I_{11010}^0&=&\int_k\frac{1+2\NB(k)}{2k} I_{11},\\
I_{12010}^0&=&\int_k\frac{1+2\NB(k)}{2 k} I_{21},\\
I_{01110}^0&=&\int_k\frac{1-2\NF(E_k)}{2E_k} I_{11},\\
I_{02110}^0&=&\int_k\frac{1-2\NF(E_k)}{2E_k} I_{21}.
\eea
The truly NLO sum-integrals are given below using the notation $\Delta_\pm=E_p\pm E_{pk}$ and $\Delta_{\sigma \tau}= k+\sigma E_p+\tau E_{pk}$. Here I only give the terms proportional to $\omega\delta(\omega)$ when the other terms contributing at $\omega>0$ can be read from~\cite{Burnier:2008ia}.
\bea
I_{10110}^{0~(\omega\sim 0)} & = & I_{11001}^{0~(\omega\sim 0)}=0,\\ 
I_{11110}^{0~(\omega\sim 0)} & = & \pi \omega\delta(\omega)\int_{k,p} \frac{\nF1'}{8 E_p^2 E_{pk}  k}
    \left\{ [1+\nB0-\nF2]\left(\frac{1}{\Delta_{-+}}+\frac{1}{\Delta_{++}}\right) \right. \nonumber \\
    && \left. -[\nB0+\nF2]\left(\frac{1}{\Delta_{--}}+\frac{1}{\Delta_{+-}}\right) \right\},\label{tps40} \\
I_{11110}^{1~(\omega\sim 0)}&=&0,\\
I_{11111}^{0~(\omega\sim 0)}&=&\pi\omega\delta(\omega)\int_{k,p}\biggl[ \frac{( k^2-E_p^2-E_{pk}^2)\nF1'\nF2'}{4 E_p^2 E_{pk}^2\Delta_{++}\Delta_{-+}\Delta_{+-}\Delta_{--}}\nn
&&+\frac{\nF1'}{2 E_{pk} E_p^2  k}\biggl\{ \frac{1}{4E_{pk}^2}\left(\frac{\Delta_{++}+E_{pk}}{\Delta_{++}^2}+\frac{\Delta_{-+}+E_{pk}}{\Delta_{-+}^2}\right)\\&&\notag
  +E_{pk} \nB0\left(\frac{1}{\Delta_{+-}^2\Delta_{++}^2}+\frac{1}{\Delta_{--}^2\Delta_{-+}^2}\right)\nn
&&  +\frac{\nF2  k}{2E_{pk}^2}\left(\frac{2\Delta_+^2-E_p^2+E_{pk}^2- k^2}{\Delta_{--}^2\Delta_{++}^2}+\frac{2\Delta_-^2-E_p^2+E_{pk}^2- k^2}{\Delta_{-+}^2\Delta_{+-}^2}\right)\biggr\}\biggr].\notag
\eea
The last sum-integrals are given for all $\omega$ for completeness (as they were not all written in a suitable form for our present purpose): 
\bea
I_{01111}^0&=&\pi\int_{p,k}\Biggl[\omega\delta(\omega)\left\{\frac{\nF1'}{4E_p^2 E_{pk}^3}(1-2\nF2)+\frac{\nF1'\nF2'}{4 E_p^2 E_{pk}^2}\right\}\label{I01111}\\ \notag&&-\left(\delta(\omega-2E_p)-\delta(\omega+2E_p)\right)\frac{[1-2\nF1][1-2\nF2]}{8 E_p^2 E_{pk}\Delta_+\Delta_-} \Biggr],\\
I_{-11111}^0&=&2 I^0_{01110}+\frac12\left(\omega^2-4M^2\right)I^0_{01111}
\eea
and
\bea
I_{12110}^{0} & = &\pi\int_{k,p}\Biggl[  \omega\delta(\omega)\Biggl(\frac{1}{16 E_p^4 E_{pk}  k}
	[\nF1'-\frac{1}{2}E_p\nF1'']\times \nonumber \\
    && \hspace{-10mm} \left\{ [1+\nB0-\nF2]\left(\frac{1}{\Delta_{-+}}+\frac{1}{\Delta_{++}}\right)
    -[\nB0+\nF2]\left(\frac{1}{\Delta_{--}}+\frac{1}{\Delta_{+-}}\right) \right\} \nonumber \\
    && \hspace{-15mm} -\frac{\nF1'}{8 E_p^2 E_{pk}  k} 
	\left\{ \frac{1+\nB0-\nF2}{\Delta_{++}\Delta_{-+}}\left(\frac{1}{\Delta_{-+}}+\frac{1}{\Delta_{++}}\right) \right. \nonumber \\
    && \left. -\frac{\nB0+\nF2}{\Delta_{+-}\Delta_{--}}\left(\frac{1}{\Delta_{--}}+\frac{1}{\Delta_{+-}}\right)\right\}\Biggr)\\ 
    &&\hspace{-18mm}+\frac{\left[\delta(\omega-2E_p)-\delta(\omega+2E_p)\right]}{32k}\notag\Biggl(\frac{E_p^2(k^2+(1-4\epsilon) E_{pk}^2-E_p^2+2M^2)-4 E_{pk}^2 p^2}{4 E_p^4 E_{pk}^3 p^2}\times\\
    &&\hspace{-15mm}(1-2\nF1)\left\{ [1+\nB0-\nF2]\left(\frac{1}{\Delta_{-+}}+\frac{1}{\Delta_{++}}\right) \nonumber -[\nB0+\nF2]\left(\frac{1}{\Delta_{--}}+\frac{1}{\Delta_{+-}}\right)\right\}
    \\&&\hspace{-15mm}\nonumber-\frac{E_p^2+E_{pk}^2-k^2-2M^2}{4 E_p^2 E_{pk}^2 p^2}\left(1-2\nF1\right)\nF2'\left[\frac{1}{\Delta_{++}}+\frac{1}{\Delta_{+-}}+\frac{1}{\Delta_{-+}}+\frac{1}{\Delta_{--}}\right]\\
    &&\hspace{-15mm}+\frac{1-2\nF1}{ E_p^3 E_{pk}}\Biggl\{\left[\frac{1}{\Delta_{++}}+\frac{1}{\Delta_{-+}}\right]\left[\frac{1}{\Delta_{++}}-\frac{1}{\Delta_{-+}}+\frac{1}{E_p}\right]\left[1+\nB0-\nF2\right]\notag\\
    &&\hspace{-15mm}+\left[\frac{1}{\Delta_{--}}+\frac{1}{\Delta_{+-}}\right]\left[\frac{1}{\Delta_{--}}-\frac{1}{\Delta_{+-}}-\frac{1}{E_p}\right]\left[\nB0+\nF2\right]\Biggr\}\notag\\\notag&&\hspace{-15mm}
-\frac{1-2\nF1}{2E_p^2E_{pk}^2}\Biggl\{\left(\frac{\Delta_+}{E_p}+\frac{E_{pk}^2-E_p^2-k^2}{2 p^2}  \right)\left[ \frac{1+\nB0-\nF2}{\Delta_{++}^2}+\frac{\nB0+\nF2}{\Delta_{--}^2} \right]\\&&\hspace{-10mm}+
\left( \frac{\Delta_-}{E_p}+\frac{E_{pk}^2-E_p^2-k^2}{2 p^2}  \right)\left[ \frac{1+\nB0-\nF2}{\Delta_{-+}^2}+\frac{\nB0+\nF2}{\Delta_{+-}^2} \right]\Biggr\}
\Biggr)\notag\\
    &&\hspace{-20mm}\notag+\left[\delta(\omega-\Delta_{++})-\delta(\omega+\Delta_{++})\right]\frac{(1+\nB0)(1-\nF1-\nF2)+\nF1\nF2}{8 E_p E_{pk} k\Delta_{++}^2\Delta_{-+}^2}
\\
    &&\hspace{-20mm}\notag+\left[\delta(\omega-\Delta_{--})-\delta(\omega+\Delta_{--})\right]\frac{-\nB0(1-\nF1-\nF2)+\nF1\nF2}{8 E_p E_{pk} k\Delta_{--}^2\Delta_{+-}^2}
\\
&&\hspace{-20mm}\notag+\left[\delta(\omega-\Delta_{+-})-\delta(\omega+\Delta_{+-})\right]\frac{\nB0\nF1-(1+\nB0)\nF2+\nF1\nF2}{8 E_p E_{pk} k\Delta_{--}^2\Delta_{+-}^2}
\\
&&\hspace{-20mm}\notag+\left[\delta(\omega-\Delta_{-+})-\delta(\omega+\Delta_{-+})\right]\frac{\nB0\nF2-(1+\nB0)\nF1+\nF1\nF2}{8 E_p E_{pk} k\Delta_{++}^2\Delta_{-+}^2}
    \Biggr].
\eea
 
\section{Full NLO result} \label{C}

To obtain the full NLO result form the previous formulas, the integrals over $\bp,\bk$ still have to be performed. From the antisymmetry of the spectral function it is enough to consider the case $\omega\geq0$. Using rotation symmetry, it is obvious that all the angular integrals are trivial up to the one containing the angle $\theta$ between $\bp$ an $\bk$ so that we have three non-trivial integrals to perform over $|\bp|, |\bk|, \cos\theta$.
The full result can be split into three parts, a part proportional to $\omega\delta(\omega)$ which we call $\rho^t_{NLO}$ for transport, a part proportional to $\delta(\omega-2E_p)$, called $\rho^f_{NLO}$ for factorized and a last part proportional to one of the $\delta(\omega\pm\Delta_{\pm\pm})$ denoted by $\rho^p_{NLO}$ for phase space, following the notation of \cite{Burnier:2008ia}. 
For the terms proportional to $\delta(\omega)$ the three integrals have to be performed but the angular integral happens to be analytically doable~\cite{Burnier:2012ze}. For the ones proportional to $\delta(\omega-2E_p)$ one can constrain the $|\bp|$ integral and only two integrals are left but most of them have to be performed numerically. The terms proportional to $\delta(\omega\pm\Delta_{\pm\pm})$ require more work as the domains where the $\omega=\pm\Delta_{\pm\pm}$ constraints can be satisfied are non-trivial.

\subsection{$\delta(\omega)$ terms}
All the terms proportional to $\delta(\omega)$ in the master integrals can be combined according to equation (\ref{NLO}). After performing an analogous work as in ref.~\cite{Burnier:2012ze}, i.e. integrating by parts and performing the angular integrals we get:
\bea
\frac{\rho^t_{NLO}}{4\Nc\CF g^2}&=&\label{rhoNLOt}\frac{ \omega \delta(\omega)}{4\pi^3}\int_0^\infty dp\, \NF'(E_p)\int_0^\infty dk\,\Biggr[ a k\,\NB(k)\left(1-\frac{3p^2}{E_p^2}\right)\\\notag &&+\frac{\NF(E_k)}{E_k}\Biggl(a k^2-M^2-a\frac{3 k^2 p^2}{E_p^2}-\frac{M^2 k^2 p^2}{E_p^2E_k^2}\\ &&-\frac{M^2 k p}{2 E_k^2E_p^2}\left\{\frac{M^2 E_k^2(a-1)}{p^2}-2(a+1)E_k^2+M^2\right\}\ln\left|\frac{p+k}{p-k}\right|\Biggr)\Biggr]\notag,
\eea
where the parameter $a$ keeps track of the thermal mass shift. Setting $a=0$ is equivalent to performing the thermal mass shift (\ref{MassShift}), otherwise $a=1$.

\subsection{$\delta(\omega-2E_p)$ terms}
Summing all terms proportional to $\delta(\omega-2E_p)$ occurring in equation (\ref{NLO}) we get a formula of the form
\bea
\rho^f_{NLO}(\omega)&=&\int_p\int_k \frac{ g(k,E_p,E_{pk},\omega)}{k E_p E_{pk}} 2 \delta(\omega-2E_p).\eea
While the function $g$ can be easily constructed from the formulas of the above section, it is very long and will not be given here. Note that this integral contains divergent terms and requires a careful renormalization. This proceeds as at zero temperature \cite{Burnier:2008ia} and will not be explained here. The thermal part we calculate here is finite. We can rewrite the previous integral as:
\bea
\rho^f_{NLO}(\omega)&=&\notag\frac{1}{8\pi^4}\int_0^\infty dk \int_M^\infty dE_p \int_{E_{pk}^-}^{E_{pk}^+}dE_{pk}\,  g(k,E_p,E_{pk},\omega) 2\delta(\omega-2E_p)\\&&=\frac{1}{8\pi^4}\int_0^\infty dk \int_{E_{pk}^-}^{E_{pk}^+}dE_{pk}\, g\left(k,\frac\omega2,E_{pk},\omega\right)\theta(\omega-2M),
\eea
where $E_{pk}^\pm=\sqrt{E_p^2+k^2\pm2 p k}\to\sqrt{\omega^2/4+k^2\pm  k \sqrt{\omega^2-4 M^2}}$ after applying the delta function. The last two integrals are performed numerically and
\beq
\rho^f_{NLO}(\omega)=\theta(\omega-2 M)\left(1-2\NF\left(\frac{\omega}{2}\right)\right)\left(\rho^{vac,f}_{NLO}+\rho^{bos,f}_{NLO}\right)+\rho^{ferm,2}_{NLO}(\omega)\label{f},
\eeq
where
\bea
&&\hspace{-5mm}\frac{\rho^{ferm,2}_{NLO}(\omega)}{4 \Nc \CF g^2}=\label{rhoNLOferm2}\\&&
+\frac{\theta(\omega-2 M)}{8 \pi^3}\left(1-2 \NF\left(\frac{\omega }{2}\right)\right)\int_0^\infty dk\int_{E_{pk}^-}^{E_{pk}^+}dE_{pk}\notag\Biggl(\nF2'\frac{ M^2 
 \left(2 M^2+\omega ^2\right) }{E_{pk} (\omega^2-4M^2) \omega }\\&&\notag\times \Biggl[k+ \frac{\left(2M^2+E_{pk} \omega\right)}{2}\left[\frac{1}{\delta_{++}}+\frac{1}{\delta_{--}}\right] + \frac{\left(2M^2-E_{pk} \omega\right)}{2}\left[\frac{1}{\delta_{+-}}+\frac{1}{\delta_{-+}}\right]\Biggr]
\\&&\notag+\nF2\Biggl[\frac{2M^2+\omega^2}{\omega E_{pk}^2(\omega^2-4 M^2)}\Biggl\{\frac{\omega M^2 E_{pk}}{2}\left(\frac{\delta_-}{\delta_{+-}^2}-\frac{\delta_-}{\delta_{-+}^2}-\frac{\delta_+}{\delta_{++}^2}+\frac{\delta_+}{\delta_{--}^2}\right)
\\&&\notag
-M^2(M^2-E_{pk}^2(1-a))\left(\frac{1}{\delta_{++}}+\frac{1}{\delta_{+-}}+\frac{1}{\delta_{-+}}+\frac{1}{\delta_{--}}\right)+k(a E_{pk}^2-M^2)\Biggr\}
\\&&\notag+\frac{1}{2\omega^2}\Biggl\{-2 k \omega a+\frac{M^2(2M^2+\omega^2)}{E_{pk}}\left(\frac{\delta_+}{\delta_{-+}^2}+\frac{\delta_-}{\delta_{++}^2}-\frac{\delta_-}{\delta_{--}^2}-\frac{\delta_+}{\delta_{+-}^2}\right)+
\\&&\notag+\left(\mathcal{P}\frac{1}{k \delta_-}+\frac{1}{k \delta_+}\right) \left(k^2 \left(4 M^2+3 \omega ^2\right)+8 M^4-2 \omega^4\right)
\\&&\notag+\left(\frac{1}{k\delta_{--}}+\frac{1}{k\delta_{-+}}\right) \left(-2 k^2 \omega ^2+\left(k
   \omega +2 M^2\right)^2-\omega ^2 (k-\omega )^2\right)
\\&&\notag+\left(\frac{1}{k\delta_{+-}}+\frac{1}{k\delta_{++}}\right) \left(2 k^2 \omega ^2-\left(2 M^2-k \omega \right)^2+\omega ^2 (k+\omega )^2\right)
\\&&\notag-2\omega(1-a)\left(\frac{1}{\delta_{++}}+\frac{1}{\delta_{+-}}+\frac{1}{\delta_{-+}}+\frac{1}{\delta_{--}}\right)\Biggr) 
\Biggr\}
\Biggr]\Biggr)
\eea
with $\delta_{\pm\pm}=2k\pm\omega\pm E_{pk}$, $\delta_\pm=\omega\pm2E_{pk}$ and again, setting $a=0$ is equivalent to performing the thermal mass shift (\ref{MassShift}).  The symbol $\mathcal{P}$ means that we treat the pole at $\delta_-=0$ in the principal value sense. Numerically this can be implemented as follows. We split the integral as $\int_{E_{pk}^-}^{\omega/2}+\int^{E_{pk}^+}_{\omega/2}$, perform a change of integration variable $E_{pk}\to \omega-E_{pk}$ in the second integral and add it to the first one.

\subsection{$\delta(\omega\pm\Delta_{\pm\pm})$ terms}
Summing all terms proportional to $\delta(\omega \pm\Delta_{\pm\pm})$ occurring in equation (\ref{NLO}) we get a formula of the form
\bea
&&\sum_{\pm,\pm}\int_p\int_k \frac{f_{\pm\pm}(k,E_p,E_{pk},\omega)}{k E_p E_{pk}} (\delta(\omega -\Delta_{\pm\pm})-\delta(\omega +\Delta_{\pm\pm})).
\eea
If we restrict ourselves to $\omega>0$, only half of the $\delta$'s can actually be realized in some domain of the integrals so that the previous sum becomes
\bea
&&\notag\frac{\theta(\omega-2M)}{8\pi^4}\int_0^{k^{1}}dk \int_{E_p^{1-}}^{E_{p}^{1+}}dE_p\,f_{++}(k,E_p,\omega-E_p-k,\omega)\\
&&\notag-\frac{1}{8\pi^4}\int_{k^{2}}^{\infty}dk \int_{E_p^{2-}}^{E_{p}^{2+}}dE_p\,f_{--}(k,E_p,\omega-E_p+k,\omega)\\
&&\notag+\frac{1}{8\pi^4}\int_{k^3}^{\infty}dk \int_{E_p^{1+}}^{\infty}dE_p\,f_{+-}(k,E_p,-\omega+E_p+k,\omega)\\
&&+\frac{1}{8\pi^4}\int_{k^3}^{\infty}dk \int_{-E_p^{1-}}^{\infty}dE_p\,f_{-+}(k,E_p,\omega+E_p-k,\omega),\label{integrated_deltas}
\eea
where the boundaries of the integrals are given by
\bea
k^{1}&=&\frac{\omega^2-4M^2}{2\omega}\notag, \quad k^{2}=\max\left(0,-k^1\right) ,\quad k^{3}=\frac\omega2,\\\notag
E_p^{1\pm}&=&\frac{\omega-k}{2}\pm\frac{k}{2}\sqrt{1-\frac{4M^2}{\omega(\omega-2k)}},\\
E_p^{2\pm}&=&\frac{\omega+k}{2}\pm\frac{k}{2}\sqrt{1-\frac{4M^2}{\omega(\omega+2k)}}
\eea
and the functions $f_{\pm,\pm}$ through
\bea 
f_{++}(k,E_p,\omega-E_p-k,\omega)&=&
\pi\Biggl(\frac{4 k M^4}{\omega ^2 (\omega -2 E_p)^2 (2 (E_p+k)-\omega )}\\&&\hspace{-45mm}+\frac{M^2 (3 \omega -4 E_p)}{\omega  (\omega -2 E_p)^2}+\frac{2 (E_p+k-\omega )^2}{(2
   E_p-\omega ) (2 (E_p+ k)-\omega )}\Biggr) \notag\\ \notag&&\hspace{-45mm}\times\left[(\nB0+1) (1-\NF(\omega-E_p-k)-\nF1)+\nF1 \NF(\omega-E_p-k )\right],\\
f_{--}(k,E_p,\omega-E_p+k,\omega)&=&
\pi\Biggl(-\frac{4 k M^4}{\omega ^2 (\omega -2 E_p)^2 (2(E_p-k)-\omega )}\\&&\hspace{-45mm}+\frac{ M^2 (3 \omega -4 E_p)}{\omega  (\omega -2 E_p)^2}+\frac{2  (k-E_p+\omega )^2}{(2
   E_p-\omega ) (2 (E_p- k)-\omega )}\Biggr)  \notag\\ \notag&&\hspace{-45mm}\times \left[\nF1 \NF(\omega-E_p+k)-\nB0 (1-\NF(\omega-E_p+k )-\nF1)\right],\\
f_{+-}(k,E_p,E_p+k-\omega,\omega)&=&
\pi\Biggl(\frac{4 k M^4}{\omega ^2 (\omega -2 E_p)^2 (2 (E_p+k)-\omega )}\\&&\hspace{-45mm}+\frac{  M^2 (3 \omega -4 E_p)}{\omega  (\omega -2 E_p)^2}+\frac{2  (E_p+k-\omega )^2}{(2
   E_p-\omega ) (2 (E_p+ k)-\omega )}\Biggr)  \notag\\ \notag&&\hspace{-45mm}\times\left[-(\nB0+1) \NF(E_p+k-\omega )+\nF1 \nB0+\nF1 \NF(E_p+k-\omega )\right]
\eea
and
\bea
f_{-+}(k,E_p,\omega+E_p-k,\omega)&=&
\pi\Biggl(-\frac{4 k M^4}{\omega ^2 (2 E_p+\omega )^2 (2 (E_p- k)+\omega )}\\&&\hspace{-45mm}+\frac{ M^2 (4 E_p+3 \omega )}{\omega  (2 E_p+\omega )^2}+\frac{2 (E_p-k+\omega )^2}{(2
   E_p+\omega ) (2 (E_p- k)+\omega )}\Biggr) \notag \\&& \notag\hspace{-45mm}\times \left[\nB0\NF(E_p-k+\omega )-\nF1 (\nB0+1)+\nF1 \NF(E_p-k+\omega )\right].
\eea
Note that at $T=0$ only the first integral in (\ref{integrated_deltas}) contributes. This term, setting $T=0$, added to the expression $\rho^{vac,f}_{NLO}$ in equation (\ref{f}) gives the full zero temperature result $\rho^{vac}_{NLO}$ given in formula (\ref{rhoNLOvac}).  For $M\gg T$ the first two integrals contribute. If we take in these terms the part proportional to $\NB(k)$ and not containing any Fermi-Dirac distribution function and add them to the expression $\rho^{bos,f}_{NLO}$ in equation (\ref{f}), we get the 'bosonic' thermal correction $\rho^{bos}_{NLO}$ given in formula (\ref{rhoNLObos}). 
The remaining terms are exponentially suppressed if $M\gg T$ but dominate the spectrum at small $\omega$.

Even if the full result is infrared finite, the different terms are not. To avoid such problems one can first add the two last integrals in (\ref{integrated_deltas}) after having performed the shift $E_p\to E_p-\omega+k$ in the last integral. As a result of that we get the three last lines of formula (\ref{rhoNLOferm}) and then in all terms add an $\epsilon$ to the lower bound of the $k$ integration and take the limit $\epsilon\to0$ after having added all terms.

\subsection{Fermionic contribution}

The full result is the sum of the vacuum part (\ref{rhoNLOvac}), the 'bosonic' thermal corrections (\ref{rhoNLObos}) and the fermionic contribution, which we can write as
\bea
\rho_{NLO}^{ferm}=\rho_{NLO}^{t}-2\NF\left(\frac{\omega}{2}\right)(\rho_{NLO}^{vac}+\rho_{NLO}^{bos})+\rho_{NLO}^{ferm,1}+\rho_{NLO}^{ferm,2}+\rho_{NLO}^{ferm,3},
\eea
where $\rho_{NLO}^{t}$ is given in formula (\ref{rhoNLOt}), $\rho_{NLO}^{vac}$ in (\ref{rhoNLOvac}), $\rho_{NLO}^{bos}$ in (\ref{rhoNLObos}), $\rho_{NLO}^{ferm,1}$ in (\ref{rhoNLOferm1}), $\rho_{NLO}^{ferm,2}$ in (\ref{rhoNLOferm2}) and the remaining $\rho_{NLO}^{frem,3}$ below:
\bea
&&\hspace{-5mm}\frac{\rho_{NLO}^{frem,3}}{4\Nc\CF g^2}=\frac{\theta(\omega-2M)}{8\pi^3}\int_0^{k^{1}}dk \int_{E_p^{1-}}^{E_{p}^{1+}}dE_p\,\Biggl(\frac{4 k M^4}{\omega ^2 (\omega -2 E_p)^2 (2 (E_p+k)-\omega )}\notag\\&&+\frac{M^2 (3 \omega -4 E_p)}{\omega  (\omega -2 E_p)^2}+\frac{2 (E_p+k-\omega )^2}{(2
   E_p-\omega ) (2 (E_p+ k)-\omega )}\Biggr) \\ \notag&&\times\left[(\nB0+1) (2\NF(\omega/2)-\NF(\omega-E_p-k)-\nF1)+\nF1 \NF(\omega-E_p-k )\right]\\
&&\notag-\frac{1}{8\pi^3}\int_{k^{2}}^{\infty}dk \int_{E_p^{2-}}^{E_{p}^{2+}}dE_p\,\Biggl(-\frac{4 k M^4}{\omega ^2 (\omega -2 E_p)^2 (2(E_p-k)-\omega )}\\&&+\frac{ M^2 (3 \omega -4 E_p)}{\omega  (\omega -2 E_p)^2}+\frac{2  (k-E_p+\omega )^2}{(2
   E_p-\omega ) (2 (E_p- k)-\omega )}\Biggr)  \notag\\ \notag&&\times \left[\nF1 \NF(\omega-E_p+k)-\nB0 (2\NF(\omega/2)-\NF(\omega-E_p+k )-\nF1)\right].
\eea

\section{Mass shift}\label{A4}
The results of ref. \cite{Burnier:2012ze} for the Euclidean correlator can be modified to account for the mass shift (\ref{MS}). We have to add to $\frac{G_V(\tau)}{4 \Nc\CF g^2}$ in equations (4.4-4.5) of \cite{Burnier:2012ze} the following integral
\bea
\frac{G^{MS}_V(\tau)}{4 \Nc\CF g^2}&=&\int_p \delta M_T^2\Biggl[\frac{p^2}{2E_p^4}\left(D_{2E_p}(\tau)+ 2 T \NF'(E_p)\right)\\&&+\left(1+\frac{M^2}{2E_p^2}\right)\frac{\partial_{E_p}D_{2E_p}(\tau)}{2E_p}+\frac{M^2}{2E_p^3}\NF''(E_p) \Biggr]\notag.
\eea
In formula (\ref{rhoNLObos}) the thermal mass shift has been performed. Without mass shift, $\frac{\rho^{bos}_{NLO}}{4\Nc\CF g^2}$ would contain an additional 
\beq
\frac{\theta(\omega-2M)}{(4\pi)^3\omega^2}\int_0^\infty dk\frac{2 \NB(k)}{k}\left(4 \omega k^2\sqrt{\omega^2-4 M^2}-4 k^2\omega\frac{(2M^2+\omega^2)}{\sqrt{\omega^2-4 M^2}}\right).
\eeq

\end{document}